\documentclass[a4paper,12pt]{article}

\makeatletter
\renewcommand\paragraph{\@startsection{paragraph}{4}{\z@}%
  {-3.25ex\@plus -1ex \@minus -.2ex}%
  {1.5ex \@plus .2ex}%
  {\normalfont\normalsize\bfseries}}
\makeatother

\makeatletter
\renewcommand\subparagraph{\@startsection{subparagraph}{4}{\z@}%
  {-3.25ex\@plus -1ex \@minus -.2ex}%
  {1.5ex \@plus .2ex}%
  {\normalfont\normalsize\bfseries}}
\makeatother

\usepackage[latin1]{inputenc}
\usepackage[english]{babel}
\usepackage{theorem}
\usepackage{amsmath}
\usepackage{amsfonts}
\usepackage{amssymb}
\usepackage{amscd}
\usepackage{amstext}
\usepackage{amsbsy}
\usepackage{amsopn}
\usepackage{amsxtra}
\usepackage{color}
\usepackage{euscript}
\usepackage{empheq}
\usepackage{upref}
\usepackage{graphicx}
\usepackage{subfig}
\usepackage{verbatim}
\usepackage[francais,verbose]{layout}
\usepackage{fancyhdr}
\usepackage{psfrag}

\theorembodyfont{ \upshape}

\theoremstyle{break} 

\newtoks\ladate
\newtoks\sujet

pt\belowdisplayshortskip 5 pt plus 3 pt minus 3 pt

\begin{document}
%

\title{Valuation of Zynga}
\author{Zal\'an Forr\'o, Peter Cauwels and Didier Sornette}
\maketitle


\section{Introduction}
\noindent After the recent initial public offerings (IPOs) of some of the major social networking companies such as Groupon, Linkedin or Pandora, Zynga went public on December 16. The estimated value of this social network game developing company was as high as 14 billion USD in November 2011 (\cite{yahoo}). However, after the poor performances of the other IPOs earlier this year, this number was scaled down. Indeed, 100 million shares of Class A common stock were sold at $10\$$ per share, the top end of the indicative $8.5\$-10\$$ range. Having a total of 699 million shares outstanding (\cite{sec}), the value of the company at IPO was of 6.9 billion USD. As of December 23, its market capitalization was of 6.6 billion USD (\cite{bloomberg}). In no case, to our knowledge, were specifics given about the methodology used to obtain these estimates, and one could wonder about the economic justification for such a change in value in only a few months period. In this paper, we extend to Zynga the methodology proposed by Cauwels and Sornette in 2011 (\cite{cauwels11}) for the valuation of Facebook and Groupon, by introducing a semi-bootsrap approach to forecast Zynga's user base.\\ \\
\noindent This paper is organized as follows. Section \ref{methodology} gives a brief summary of the methodology used to value Zynga. Section \ref{users} describes the dynamics of the number of daily active users (DAU) of Zynga. Section \ref{financials} analyzes the financial informations relevant to the valuation of the company. Section \ref{valuation} gives its estimated market capitalization and section \ref{conclusion} concludes.

\section{ Valuation methodology} \label{methodology}
The major part of the revenues of a social networking company is inherently linked to its user base. The more users it has, the more income it can generate through advertising. The basic idea of the method, proposed by Cauwels and Sornette in 2011 (\cite{cauwels11}), is to separate the problem into 3 parts:
\begin{enumerate}
\item A part based on what we consider as hard data: this part uses the known historic daily active users (DAU) metric that quantifies the frequentation of Zynga. Based on this figure, we develop a methodology to forecast the future evolution of Zynga's user base. Other metrics exist such as the number of registered users (U) in the case of Facebook. To understand the reason why the DAU instead of the U are used for Zynga, we need to explain how this company is organized. Zynga uses Facebook as a platform for its games. One does not need to register in order to get access to Zynga's games; it is enough to have a Facebook account. Therefore, the number of registered users is not even well defined for Zynga. The choice of the DAU metric has consequences: being a more dynamical measure than U it can decrease significantly from one day to another. This is not true for the number of registered users: in practice, it takes an effort in order to unregister from Facebook. Because U are monotonically increasing, the logistic model (equation \ref{logistic}) could be used by Cauwels and Sornette (\cite{cauwels11}) to model the growth dynamic.
\begin{equation} \label{logistic}
\frac{dU}{dt} = rU(1 - \frac{U}{K})
\end{equation}

\noindent Here $r$ is the constant growth rate and $K$ the carrying capacity (biggest possible number of users).\\
\noindent For Zynga however, a different approach had to be worked out. Another specificity of Zynga, justifying the need for a new approach, resides in the fact that it is a heterogeneous entity: it is the aggregate of all the games it developed. Therefore, to understand the dynamics of Zynga's user base, we had to examine the dynamics of the DAU of the individual games it developed (figure \ref{dau_time}). 

\begin{figure}[!h] 
    \centering
    \psfrag{1}[B][B][1.2][0]{s(t)}
    \includegraphics[width=1\textwidth]{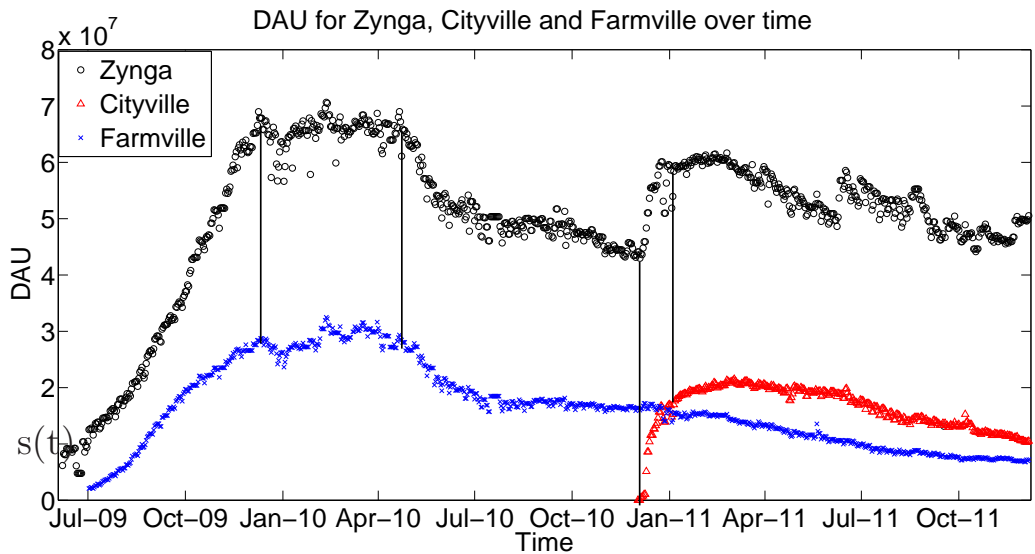}
    \caption{Number of DAU in time for Zynga and its 2 most popular games (Cityville and Farmville; only these 2 games are reported for clarity). A typical feature of the games can be seen in Cityville and Farmville: after an initial rapid raise, the DAU of the games enters a decay phase. The black vertical lines illustrate the notion that the DAU of Zynga depends on the performance of its individual games. Source of the data: \emph{http://www.appdata.com/devs/10-zynga}}
    \label{dau_time}
\end{figure}

\noindent In order to reproduce the DAU of Zynga, only the top 20 games (out of 63) were considered; these account for more than $97\%$ of the total number of Zynga users. The specifics about the dynamics of Zynga and its individual games will be further elaborated in section \ref{users}.
 
\item Another part of the methodology is based on what we consider as soft data: this part uses the financial data available in the S1 form of  filings to the SEC (\cite{sec}). The data are used to give us an estimate of how much revenues are generated per daily active user over time. It also gives us information about the profitability of the company. Due to the limited amount of published financial information, we also have to rely on our intuition and good-sense to give our best forecast of the future revenues per user generated by the company. This part will be further elaborated in section \ref{financials}.

\item The third part combines the two previous parts to value the company: having an estimate of the future number of users (DAU), and of the revenues each of them will generate ($r$), it is possible to compute the future revenues of the company. The revenues are converted into profits using a best-estimate profit margin ($p_{\text{margin}}$) and are discounted using the equity risk premium $d$. The net present value of the company is then the sum of the discounted future profits (or cash flow).

\begin{equation} \label{eq:valuation}
Valuation = \sum_{t=1}^{end}  \frac{r(t) \cdot DAU(t) \cdot p_{\text{margin}}}{(1+d)^t} = \sum_{t=1}^{end} \frac{\text{profits}}{(1+d)^t}
\end{equation}

\noindent Hereby we optimistically assume that all the profit is distributed to the shareholders.

\end{enumerate}

\section{Hard Data} \label{users}

\subsection{General approach}
To predict the evolution of the user base of Zynga, the following steps were taken. 
\begin{enumerate}
\item We used the functional form of each of the top 20 games to forecast the future DAU evolution of the company. This was done as follows:
	\begin{itemize}
	\item For the data that are available, we did not use any functional form; we used the actual data as the best proxy for the DAU.
	\item To extend the DAU dynamics into the future, we fitted a functional form to the decay process. 
	\end{itemize}
\item Relying heavily on the creation of new games in order to maintain/increase its DAU (figure \ref{dau_time}), it was important to quantify the rate of innovation of Zynga. This was defined as $\tau$, the average time between the implementation of new games (restricted to the top 20).
\item Finally, a future scenario could be simulated as follows: each $\tau$ days for the next 20 years (after present), a game was randomly chosen among the top 20. Its dynamics were calculated according to the first step. The DAU of Zynga over time was then simply the sum of all the simulated games. A 1000 different scenarios were computed.
\end{enumerate}

\subsection{The tails of the DAU decay process}
\begin{figure}[!h]
    \centering
    \includegraphics[width=1 \textwidth]{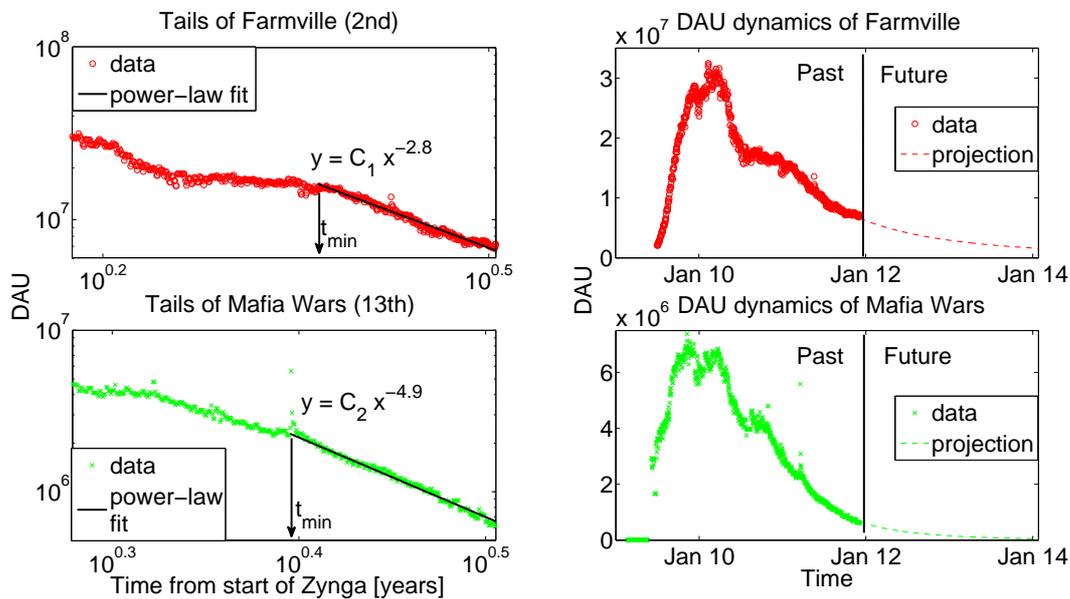}
    \caption{Left: Decay process of Farmville (red circles) and Mafia Wars (green crosses), the 2nd and 13th most popular games respectively. We can see that a power-law is a  good fit for the tails from $t_{min}$. Right: Simulated dynamics of the games based on the power-law parameters of panel a). To forecast the DAU of a given game, the fitting parameters of the functional form (power-law) for the available time span were used. Source of the data: \emph{http://www.appdata.com/devs/10-zynga}}
    \label{tails}
\end{figure}

The functional form of the DAU of each game is composed of the actual observed data and a tail that simulates the future decay process. For the majority of games, it was found that a power-law, $f(t) \propto t^{-\gamma}$, is a reasonable approximation (figure \ref{tails}). This choice was also motivated by the slower decay of the power-law compared to the exponential one, and as such prevents an unnecessary devaluation of the company.\\

 \subsection{Predicting the future DAU of Zynga}
\begin{figure}[!h] 
    \centering
    \includegraphics[width=1\textwidth]{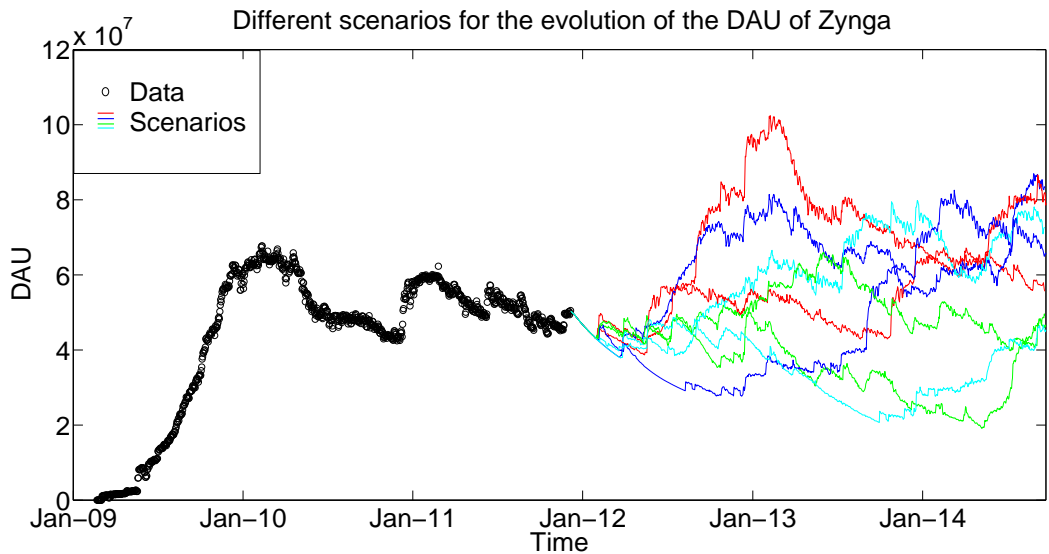}
    \caption{4 different scenarios of the DAU evolution of Zynga for the first 3 years. The company will be valued for each of these scenarios (section \ref{valuation}), giving a range for the its expected market capitalization.}
    \label{scenarios}
\end{figure}

The average time between the implementation of new games from the top 20 was found to be $\tau \approx 53$ days. A game belonging to the top 20 was randomly chosen each $\tau$ days for the next 20 years (starting from the present). Summing the DAU of all these games, the user's dynamics of Zynga was computed. This process was repeated 1000 times, giving 1000 different scenarios. As can be seen from figure \ref{scenarios}, the evolution of the user base between scenarios can be quite different. That is the reason why a wide range of scenarios are needed. The valuation of the company will be computed for each of those scenarios (from equation \ref{eq:valuation}), giving a probabilistic forecast about the market capitalization of Zynga. 

\section{Soft Data} \label{financials}
The financial information concerning Zynga can be found in the Form S1 filing to the SEC (\cite{sec}). The crucial information for our purpose was the revenues per DAU per year as explained in section \ref{methodology}. First, the yearly revenues were derived from the quarterly as follows:
\[
R_i = R^q_{i-3} + R^q_{i-2} + R^q_{i-1} + R^q_{i}  
\]

\begin{figure}[!h] 
    \centering
    \includegraphics[width=1\textwidth]{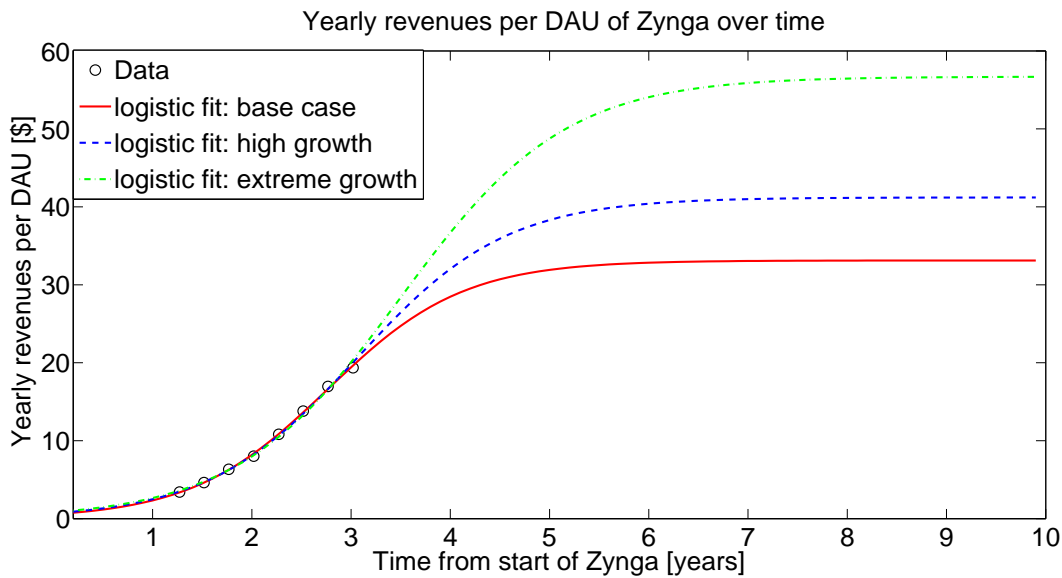}
    \caption{Yearly revenue per DAU, $r_i$ over time. A logistic fit (equation \ref{logistic}) is proposed with $K \approx 33, 41$ and $57$ USD/DAU for the base case, high and extreme growth case scenarios. The variable $P$ of equation \ref{logistic} is thus $r_i$ in this case. Source of the data: S1 form of filings to the SEC (\cite{sec}).}
    \label{rev_per_dau}
\end{figure}

\noindent Here $R_i$ and $R^q_i$ are respectively the yearly and quarterly revenues at quarter $i$ with $i \in (4, \text{last})$. The yearly revenues per DAU at each quarter, $r_i$, were then obtained by dividing $R_i$ by $\langle DAU_i \rangle_{year}$, the realized DAU at time $i$ averaged over the preceding year (figure \ref{rev_per_dau}). Using the methodology proposed by Cauwels and Sornette (\cite{cauwels11}) for the hard data, logistic growth functions are fitted to the observations (figure \ref{rev_per_dau}) representing respectively a base case, high growth and an extreme growth scenario for the soft data. For the high growth and extreme growth scenarios, $K$ is fixed at its $80\%$ and $95\%$ confidence value respectively during the fitting (\cite{cauwels11}).

\noindent It is not surprising to see a saturation in the revenues/user as indicated by the logistic fit. In fact, there have to be constraints on how much money can be extracted from a user. Indeed, under the spacial constraints (there is a limited number of advertisements that can be displayed on a webpage), time constraints (there are only so many advertisements that can be shown one day) and ultimately the economical constraints (there is only so much money a user can spend on games or as an advertiser is willing to spend), the revenues per DAU are doomed to saturate. Using this logistic description for the revenues per DAU, a valuation of Zynga will be given for each of the 3 growth hypothesis.

\section{Valuation} \label{valuation}
Combining,  through equation \ref{eq:valuation}, the hard part of the analysis with the soft part, i.e. the number of users over time and the revenues each of them generates per year, the value of the company can be revealed.

\begin{figure}[!h] 
    \centering
    \includegraphics[width=1\textwidth]{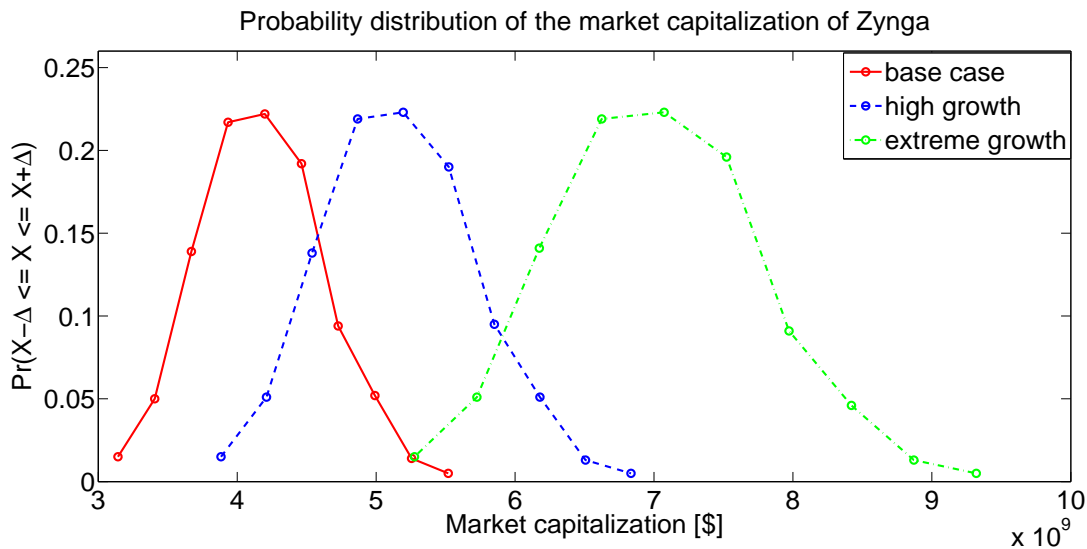}
    \caption{Market capitalization of Zynga according to the base growth, high growth and extreme growth scenarios. This shows that the 7B USD valuation is only satisfied in the extreme revenues case.}
    \label{fig:valuation}
\end{figure}

\noindent We used an optimistic $15\%$ profit margin; the profit margin of last fiscal year (and the highest so far), although it was a little below $4 \%$ for the first 3 quarters of this fiscal year (\cite{sec}). We also assumed that all profits will be distributed to the shareholders and used a discount factor of $5 \%$ as in \cite{cauwels11}. Then we computed the company's valuation for all 1000 different scenarios using equation \ref{eq:valuation}. The results are shown in figure \ref{fig:valuation} and table \ref{table}.

\begin{table} [!h]
    \centering
    \begin{tabular}{ | l | l | l |}
    \hline
    Scenario [billion USD] & Valuation & $95\%$ two-sided confidence interval \\ \hline
    Base case & 4.17 & [3.36; 5.08]  \\ \hline
    High growth & 5.16 & [4.15; 6.29]  \\ \hline
    Extreme growth & 7.02 & [5.63; 8.56]  \\ \hline
    \end{tabular}
    \caption{Valuation of Zynga in the base case, high and extreme growth scenarios.}
    \label{table}
\end{table}

\noindent We obtain a valuation of 4.17B USD for our base case scenario, well below the $\approx$ 7B USD value at IPO. Only the unlikely extreme growth case scenario could potentially justify Zynga's valuation at IPO.

\section{Conclusion} \label{conclusion}

\noindent In this paper, we propose a new valuation methodology to price Zynga. Our first major result was to model the future evolution of Zynga's DAU using a semi-bootstrap approach: combining the empirical data (for the available time span) together with a functional form for the decay process (for the future time span). Indeed, the user growth dynamics could not be modeled with the logistic growth function (as it was done for Facebook and Groupon (\cite{cauwels11})). This was partly due to the fact, that the evolution of the DAU in time was governed by the  specific dynamics of the individual games. \\

\noindent The second major result was that the evolution of the revenues per user in time showed a slowing of the growth rate, which we modeled with a logistic function. This makes intuitive sense as these $r_i$ should be bounded due to various constraints, the hard constraint being the economic one, since Zynga's players only have a finite wealth. We studied 3 different cases for this upper bound: the most probable one (the base case scenario), an optimistic one (the high growth scenario) and an extremely optimistic one (the extreme growth scenario).\\

\noindent On the basis of our results, we can claim with confidence that at its IPO, Zynga has been overvalued. Indeed, only the extreme growth scenario (implying 57 USD/DAU at saturation) would be able to justify the value at IPO of the company. This scenario is unlikely ($\approx 5\%$ probability) and even more so given our optimistic approach:
\begin{itemize}
\item We have taken a power-law for the decay process (even in cases where exponentials might be better).
\item We chose games only in the top 20 with equal probability in the simulation process (implying that there is the same probability to create a top game and an average/bad one).
\item We took $15 \%$ profit margin and supposed that all the future profits would be distributed to shareholders.
\item We implicitly assumed that the real interest rates and the equity risk premium stay constant at $0\%$ and $5\%$ respectively, for the next 20 years (\cite{cauwels11}).
\end{itemize}

\noindent Given these optimistic assumptions, all our estimates should be regarded as an upper bound in our valuation of Zynga.

\end{document}